\documentclass[aps,pra,showpacs,twocolumn,amsmath,amssymb,superscriptaddress,footinbib]{revtex4}

\usepackage[english]{babel}
\usepackage{latexsym}
\usepackage{graphicx}
\usepackage{subfigure}
\usepackage{epsfig}
\usepackage{amsfonts}
\usepackage{amssymb}
\usepackage{amsmath}
\usepackage{bbm}

\begin{document}

\title{Interaction induced Landau-Zener transitions}


\author{Jonas Larson}
\affiliation{Department of Physics,
Stockholm University, AlbaNova University Center, Se-106 91 Stockholm,
Sweden}
\affiliation{Institut f\"{u}r Theoretische Physik, Universit\"{a}t zu
  K\"{o}ln, De-50937 K\"{o}ln, Germany}

\date{\today}

\begin{abstract}
By considering a quantum critical Lipkin-Meshkov-Glick model we analyze a new type of Landau-Zener transitions where the population transfer is mediated by interaction rather than from a direct diabatic coupling. For this scenario, at a mean-field level the dynamics is greatly influenced by quantum interferences. In particular, regardless of how slow the Landau-Zener sweep is, for certain parameters almost no population transfer occurs, which is in stark contrast to the regular Landau-Zener model. For moderate system sizes, this counterintuitive mean-field behaviour is not duplicated in the quantum case. This can be attributed quantum fluctuations and the fact that multi-level Landau-Zener-St\"uckelberg interferences have a `dephasing' effect on the above mentioned phenomenon. We also find a discrepancy between the quantum and mean-field models in terms of how the transfer probabilities scale with the sweep velocity. 
\end{abstract}

\maketitle

\section{Introduction}
The Landau-Zener (LZ) formula gives the transition probability when a system is swept through an avoided crossing~\cite{lz,zener}. Explicitly, by introducing the {\it diabatic states} $|1\rangle$ and $|2\rangle$ and write a general state as $|\psi(t)\rangle=c_1(t)|1\rangle+c_2(t)|2\rangle$, the LZ problem solves the coupled equations  ($\hbar=1$)
\begin{equation}
i\frac{\partial}{\partial t}\left[
\begin{array}{c}
c_1(t)\\
c_2(t)\end{array}\right]=\left[
\begin{array}{cc}
\lambda t & U \\
U & -\lambda t\end{array}\right]\left[
\begin{array}{c}
c_1(t)\\
c_2(t)\end{array}\right],
\end{equation}
where $\lambda$ is the sweep velocity and $U$ the coupling strength of the two diabatic states. For an initial state $|\psi(-\infty)\rangle=|1\rangle$, the probability for population transfer from the state $|1\rangle$ to the state $|2\rangle$ at $t=+\infty$ is $P_\mathrm{LZ}=\exp\left(-\Lambda\right)$ with the {\it adiabaticity parameter} $\Lambda=\frac{2\pi U^2}{\lambda}$. In the {\it adiabatic regime}, $\Lambda\ll1$, we obtain an almost complete transfer of population between the two states. This LZ formula holds only for initial conditions as the one above (or equivalently $|\psi(-\infty)\rangle=|2\rangle$) and for infinite integration times $t\in[-T,T]$; $T\rightarrow\infty$. For finite times or other initial conditions, quantum interference alters the exponential transition formula. As will be discussed in the present work, this phenomenon is especially evident in certain non-linear extensions of the above paradigm LZ model. 

Various generalizations of the LZ problem have been considered in the past, especially multi-level problems~\cite{threelz1,bowtie,julienne}, many-body situations~\cite{altland,fleischhauer,lmgcrit,orso,LZoptlat}, and non-linear LZ transitions~\cite{nlz,wu,lmglz,lmg2}. It has been particularly demonstrated that for non-linear models both the exponential dependence and the smoothness of $P_\mathrm{LZ}$ can be lost due to {\it hysteresis} phenomena~\cite{nlz,wu}. Furthermore, in the adiabatic regime when $P_\mathrm{LZ}$ is smooth, the transition probability typically obeys a power-law dependence, {\it i.e.} $P_\mathrm{LZ}\sim\lambda^\nu$ for some exponent $\nu$~\cite{wu}. Such non-linear LZ problems arise in mean-field theories of quantum many-body problems~\cite{nlz,wu,lmglz,lmg2}. Using classical adiabaticity arguments, power-law dependences have also been predicted in many-body LZ problems beyond the mean-field regime~\cite{altland,fleischhauer}. All these works assume infinite integration times, or more precisely choosing an initial state $|\psi(-\infty)\rangle=|1\rangle$ (or the ground-state in the many-body/level setting). At these infinite initial times the diabatic and {\it adiabatic states} coincide and as a result, effects deriving from the interference phenomenon mentioned above will be greatly suppressed. It is not clear, however, how other more general initial states will evolve for non-linear models.

We note that in the above extended LZ models the transition is maintained by a constant coupling between the diabatic states. Thus, interaction in these models primarily adds an effective (non-linear) energy shift of the instantaneous ({\it adiabatic}) energies. In this work we consider a different scenario where the coupling is solely driven by interaction, such that turning off the interaction implies a trivial decoupled system. In particular, we analyze a {\it Lipkin-Meshkov-Glick model} (LMG)~\cite{lmg}, both at a mean-field and at a many-body level. At the mean-field level, by considering initial states as those discussed above ($|1\rangle$ or $|2\rangle$) they are decoupled and we encounter no population transfer. As a result, to stimulate any population transfer both initial, diabatic or adiabatic, states have to be populated and interferences between the two is unavoidable. In addition, in this ``interaction induced LZ model'', as will be shown, this type of interference has far more drastic influence on the dynamics than in the other LZ models. Beyond mean-field, at a full many-body level, quantum fluctuations will, however, act as a sort of `dephasing' and the interference phenomenon is not equally transparent. Like in other extended LZ model, both at the mean-field and the full quantum level we find a power-law dependence on the transition probability, but the exponentials differ in the two cases for the system sizes considered.

\section{Landau-Zener transitions}
Due to the diverging adiabatic energies of the LZ model in the asymptotic time limits, whenever more general initial states of the LZ problem are studied one encounters a mathematical controversy regarding quantum interferences. Let us briefly mention this by looking at the general solution of the LZ problem which can be expressed in terms of a scattering matrix;
\begin{equation}\label{smatrix}
\left[\begin{array}{c}
c_1(+\infty)\\
c_2(+\infty)\end{array}\right]=
\left[\begin{array}{cc}
S_1 & S_2\\
-S_2 & S_1^*\end{array}\right]
\left[\begin{array}{c}
c_1(-\infty)\\
c_2(-\infty)\end{array}\right].
\end{equation} 
The matrix elements are~\cite{smel}
\begin{equation}
\begin{array}{l}
S_1=\sqrt{1-P_\mathrm{LZ}}e^{i\chi},\\ \\
S_2=P_\mathrm{LZ},
\end{array}
\end{equation}
with the phase
\begin{equation}\label{fas1}
\chi=\frac{3\pi}{4}-\arg\left[\Gamma\left(i\frac{\Lambda}{2}\right)\right]+2\Phi,
\end{equation}
where the last term is related to the (adiabatic) dynamical phase accumulated throughout the transition,
\begin{equation}\label{fas2}
\Phi=\lim_{t\rightarrow\infty}\left[\frac{\lambda}{2}t^2+\frac{\Lambda}{2}\log\left(\sqrt{2\lambda}t\right)\right]
\end{equation}
and $\Gamma(x)$ is the {\it gamma function}. Obtaining the asymptotic solution above implies studying the limits of functions when their arguments $|z|$ goes to infinity. These limits may depend on the phase of $z$, something referred to as the {\it Stokes phenomenon}~\cite{stokes}. The lines in the complex plane where the function changes character are called {\it Stokes lines} and in particular for the LZ problem the $t\rightarrow-\infty$ and the $t\rightarrow+\infty$ limits belong to different sectors divided by two such Stokes lines~\cite{kalle}. 

Returning to the expressions (\ref{fas1}) and (\ref{fas2}) we have that $\Phi$ diverges in the large time limit, which means that the probability to find the system in, say, state $|1\rangle$ for an initial state $|\psi(-\infty)\rangle=\cos\theta|1\rangle+\sin\theta|2\rangle$,
\begin{equation}\label{prob}
\begin{array}{lll}
P_1 & = & \cos^2\theta\left(1-P_{LZ}\right)+\sin^2\theta P_{LZ}\\ \\
& & +\sin2\theta P_{LZ}\sqrt{1-P_{LZ}}\cos\chi,
\end{array}
\end{equation}
is ill-defined. Naturally, this is a result of looking at the asymptotic solution of the LZ problem, while for finite time sweeps the dynamical phase $\Phi$ is finite. This interference effect is well known from the theory of {\it Landau-Zener-St\"uckelberg interferometry}~\cite{stuck}: the LZ transition depends on the relative phase of the incoming state. 

It should be clear that whenever the state is initialized in say $|1\rangle$ and the initial time is negative and finite the transition probability will always display some non-monotonic behaviour due to interference between the corresponding adiabatic states. As will be demonstrated in the next section, for the interaction induced LZ problem discussed in this work, the influences from this type of interference is greatly enhanced.

\section{Interaction induced Landau-Zener transitions}
The LMG model was first introduced in nuclear physics~\cite{lmg}, but have since then been shown to be of relevance for numerous other systems, including atomic condensates in double-well traps~\cite{lmg1,lmglz}, in ion traps~\cite{lmg2}, or in cavity/circuit QED~\cite{lmg3}. The LMG model can also be seen as an {\it infinite range transverse Ising model} where every spin interact equally with each other. The type of LMG model we analyze is given by
\begin{equation}\label{lmgham}
\hat{H}_\mathrm{LMG}=\lambda t\hat{S}_z-\frac{U}{\mathcal{S}}\hat{S}_x^2.
\end{equation}
Here, $\hat{S}_x$, $\hat{S}_y$ and $\hat{S}_z$ are the $SU(2)$ angular momentum operators obeying the commutation relations $[\hat{S}_\alpha,\hat{S}_\beta]=i\varepsilon_{\alpha\beta\gamma}\hat{S}_\gamma$ with $\varepsilon_{\alpha\beta\gamma}$ the fully antisymmetric {\it Levi-Civita tensor}. The LZ sweep velocity $\lambda$ is taken to be positive, and $U$, the interaction strength, is also positive meaning that we consider the ferromagnetic case. The ``classical limit'' accounts to take the spin $\mathcal{S}\rightarrow\infty$. The diabatic states are the eigenstates of the $z$-spin component, $\hat{S}_z|\mathcal{S},m\rangle_z=m|\mathcal{S},m\rangle_z$. Importantly, we note that it is the interaction term causing a coupling between these diabatic states. With the {\it Schwinger's spin-boson mapping}~\cite{sakurai}; $\hat{S}_z=\left(\hat{a}^\dagger\hat{a}-\hat{b}^\dagger\hat{b}\right)/2$, $\hat{S}^+=\hat{a}^\dagger\hat{b}$, and $\hat{S}^-=\hat{b}^\dagger\hat{a}$, it follows that in the boson representation the interaction scatters two `$a$'-particles into two `$b$'-particles or vice versa. In addition to the continuous $U(1)$ symmetry arising from conserved spin, $\left[\hat{{\bf S}}^2,\hat{H}_\mathrm{LMG}\right]=0$, the model also supports a $\mathbb{Z}_2$ parity symmetry given by $\left(\hat{S}_x,\hat{S}_y,\hat{S}_z\right)\rightarrow\left(-\hat{S}_x,-\hat{S}_y,\hat{S}_z\right)$. If $\mathcal{S}$ is an integer, the ground state at $t=-\infty$ and at $t=\infty$ has the same parity, while this parity changes when $\mathcal{S}$ is an half integer. In the following we will always assume the spin to be an integer such that the instantaneous ground state parity is preserved through the sweep.

Thinking of $t$ as a parameter, for large $|\lambda t|$ the ground state is ferromagnetic; $|\mathcal{S},\mathcal{S}\rangle_z$. For $\lambda t=0$ instead, the ground state $|\mathcal{S},\pm\mathcal{S}\rangle_x$ is doubly degenerate. In the thermodynamic limit (here equivalent to the classical limit $\mathcal{S}\rightarrow\infty$), the model is {\it quantum critical}~\cite{sachdev} with critical points at $\lambda t/U=\pm2$. The transitions are of the {\it Ising universality class} and for $|\lambda t/U|<2$ the system is in the symmetry broken phase in which the $\mathbb{Z}_2$ parity is broken. The antiferromagnetic LMG model~(\ref{lmgham}), {\it i.e.} $U<0$, is not critical but instead there is a first order transition at $\lambda t=0$ separating the two ferromagnetic states $|\mathcal{S},\pm\mathcal{S}\rangle_z$.  

As a final remark, we compare the present LMG model to the otherwise frequently analyzed LMG systems, see Refs.~\cite{lmg1,lmglz,lmg2,lmg3}. In all these cases, a term $\epsilon\hat{S}_x$ is included in the Hamiltonian. Such a term breaks the $\mathbb{Z}_2$ symmetry and thereby split the ground state degeneracy and the model is no longer quantum critical. Equally important, the LZ transition occurs also for zero interaction $U=0$ in such cases. We call these models for {\it parity-broken LMG} systems. 


\subsection{Mean-field analysis}
As the spin is preserved, the phase space is the $SU(2)$ {\it Bloch sphere} with radius $\mathcal{S}$. The classical, or mean-field Hamiltonian, depends therefor on the polar and azimuthal angles $\theta$ and $\phi$. The corresponding classical Hamiltonian
\begin{equation}\label{clasham}
\frac{H_\mathrm{cl}}{\mathcal{S}}=\lambda t\cos(\theta)-U\sin^2(\theta)\cos^2(\phi),
\end{equation}
gives the classical equations of motion
\begin{equation}\label{claseom}
\begin{array}{l}
\dot{\phi}=\lambda t\sin(\theta)+U\sin(2\theta)\cos^2(\phi),\\ \\
\dot{\theta}=U\sin^2(\theta)\sin(2\phi),
\end{array}
\end{equation}
with the dot representing the time-derivative. Note that the above mean-field equations are ``exact'' when the quantum state is enforced to populate a spin-coherent state $|\theta,\phi\rangle$, and in particular one would expect the accuracy of this approach to be good for spins $\mathcal{S}\gg1$. Initially, $t_i=-T$, we assume the {\it magnetization} $z\equiv\cos(\theta)\approx1$. Thus, the spin precesses around the north pole. This marks an important difference between the present model and previously studied ones; if we let $z\equiv1$ we see that at a mean-field level the dynamics is frozen, {\it i.e.} no population transfer takes place. This derives from the fact that the transitions are emerging from interaction and when the ``target mode'' is empty there are no (quantum) fluctuations stimulating a transition. Thereby, we automatically have to initialize $z\neq\pm1$ and as a result the population transfer will depend on the above discussed quantum interference occurring between the adiabatic states. Note that this is regardless of integration time - also in the limit $T\rightarrow\infty$. This is very different from other non-linear LZ models where the interaction acts as an effective energy shift rather than a coupling of diabatic states~\cite{nlz,wu}. Thus, we expect the LZ interference effect to be particularly pronounced in the present model. 

\begin{figure}[h]
\centerline{\includegraphics[width=8cm]{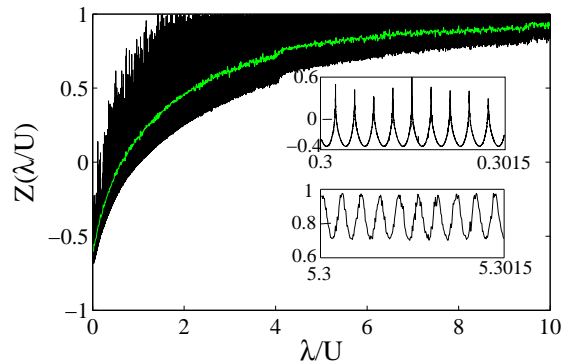}}
\caption{The imbalance $z$ for large times $t_f=200$ as a function of $\lambda/U$ (we actually average $z(t)$ over some periods around $t_f$ in order to avoid additional fluctuations). The initial condition is taken as $z(t_i)=\cos\theta=0.98$ and $\phi(t_i)=0$. The insets display zooms of the imbalance around the sweep velocity $\lambda/U\approx0.3$ and $\lambda/U\approx5$. The green solid line is the result from a TWA simulation with 5 $\%$ fluctuations in the initial imbalance $z(t_i)$ and fully random initial phase $\phi(t_i)$. } \label{fig2}
\end{figure}

In the following we will integrate the classical equations of motion from $t_i=-200$ to $t_f=+200$. The initial magnetization $z_{t_i}=0.98$, {\it i.e.} the initial spin is very close to pointing to the north pole. The magnetization after the LZ sweep, $z_{t_f}$, as a function of the LZ sweep velocity is displayed in Fig.~\ref{fig2}. In the adiabatic regime, typically $\lambda<1$, the lower bound of $z_{t_f}$ follows a power-law behaviour $\sim\lambda^\nu$ with $\nu=1/2$. The smallest $\lambda$ in the figure is $\lambda=0.02$ meaning that for $t_i=-200$ the lowest adiabatic state is predominantly populated. The LZ type interferences are evident throughout the parameter regime in terms of rapid oscillations where the transition is greatly suppressed. It is important to appreciate that the amplitude of these oscillations are much larger than what could be expected from the expression~(\ref{prob}); with $z=\cos\theta=0.98$ the amplitude of the LZ oscillations $\sin(2\theta)P_\mathrm{LZ}\sqrt{1-P_\mathrm{LZ}}<0.15$. We have numerically integrated the regular LZ problem with the same initial state and over the same time interval and found that the oscillations are often an order of magnitude smaller in amplitude in the regular LZ model. Furthermore, when the integration time is increased in the present model, the amplitude of oscillations grows and in the limit $t_i\rightarrow-\infty$ our numerical results suggest that the (adiabatic) transfer can be largely suppressed also for infinitely small $\lambda$'s. This is in stark contrast to the analytical result (\ref{prob}) for the regular LZ problem.   

Let us look closer to the behaviour of Fig.~\ref{fig2} and especially how the LZ interference can be understood in this classical picture. For large negative $\lambda t$, the azimuthal angle $\phi$ oscillates rapidly while the polar angle $\theta$ evolves on a much longer time-scale ({\it adiabatic regime}). Put in other words, whenever $\lambda|t|\gg U$, which warrants adiabatic evolution, the classical action $I=\int_0^{2\pi} z\,d\phi$, with the integration curve along the classical phase space trajectory, stays constant (equivalently, during one classical orbit the Hamiltonian change is minimal)~\cite{ll}. In the vicinity of the crossing, $\lambda t\sim0$, there is, however, no clear separation of time-scales between the two variables and it is here that adiabaticity breaks down (also called {\it sudden} or {\it critical regime}). In the limit of adiabatic evolution, the state follows the instantaneous constant energy curves, $H_\mathrm{cl}[\theta,\phi,t]=\mathrm{constant}$. The extrema of the Hamiltonian functional give the fixed points of Eq.~(\ref{claseom}). The north and south pole on the Bloch sphere are two {\it hyperbolic fixed points} of the classical equations of motion. There are two additional ({\it elliptic}) fixed points; $(\theta_\mathrm {fp},\phi_\mathrm{fp})=\left(\arccos\left(-\lambda t/2U\right)\right),0)$ and $(\theta_\mathrm{fp},\phi_\mathrm{fp})=\left(\arccos\left(-\lambda t/2U\right)\right),\pi)$. For large times $|t|$ these coincide with the other two fixed points. For $|\lambda t|/2U\leq1$, however, they traverse the Bloch sphere along the meridians $\phi=0,\,\pi$. The four fixed points defines the (non-linear) {\it adiabatic energy curves} via $E_\mathrm{ad}(t)=H_\mathrm{cl}[\theta_\mathrm{fp},\phi_\mathrm{fp}]$, which with the above expressions become $E_\mathrm{ad}(t)=\pm\mathcal{S}\lambda t$ and $E_\mathrm{ad}(t)=-\mathcal{S}\left(U+\frac{\lambda^2t^2}{4U}\right)$.

Historically, the rapid changes in the transition probability for non-linear LZ problems has been traced back to a hysteresis effect (the adiabatic energies build up so called {\it swallow-tail loops}) which is present above some critical strength of non-linearity~\cite{wu}. Dynamically, this is explained from two fixed points `colliding' in phase space and the solution is not able to precess around a single fixed point any longer. The interferences of Fig.~\ref{fig2} can also be understood by returning to the phase space evolution. Initially, the system adiabatically encircles the north pole. At some instant, in the terminology of a {\it transcritical bifurcation}, two elliptic fixed points begin to depart from the north pole. This is the critical regime where there exist no clear separation of time scales. The solution can here `chose' between encircling the hyperbolic or elliptic fixed point as they separate in phase space. In the latter case, the system ends up with a large fraction of population centered around the south pole. Thus, the rapid variations in the population transfer again stems from a `collision' of fixed points, but this time it coincides with a critical point in the original quantum model. This is indeed the crucial difference between this model and the earlier studies; the fate of the system which is determined from which fixed point it will `follow' occurs in the critical regime while in other models the evolution can be smooth up to the `hysteresis jump'. Another way to see the difference is to note that for the parity-broken LMG system, the bifurcation is of the {\it imperfect} type. To make the picture more clear, the adiabatic energy curves, showing the transcritical bifurcation, are depicted in Fig.~\ref{fig3} as blue lines (solid lines are the stable and dashed lines the unstable classical solutions). Breaking of the parity symmetry implies opening up a gap between the two elliptic solutions. In fact, it has been shown that for the ferromagnetic ($U>0$) parity-broken LMG model a non-zero interaction $U$ increases the population transfer~\cite{nlzexp} contrary to the LMG model analyzed in this work. We see that the present model also displays swallow-tail loops, but contrary to earlier studies these additional solutions are always present for non-zero $U$ and the fixed point `collision' therefor occurs as long as $U\neq0$. 

A most relevant question is if the interferences survive in the quantum case where quantum fluctuations could destabilize the classical solutions. To explore the influence of quantum fluctuations of the initial states we apply the {\it truncated Wigner approximation} (TWA)~\cite{twa} which solves the classical equations of motion for a set of initial states $(\phi_n(t_i),\theta_n(t_i))$ which are taken randomly according to the initial quantum distribution $|\Psi(\theta,\phi,t_i)|^2$. The resulting semi-classical results are obtained by averaging over the set of classical solutions, {\it i.e.} the trajectories are added incoherently meaning that any dynamical quantum interference effects are neglected. The results of a TWA simulation is presented as the green line in Fig.~\ref{fig2}. Expectedly, the initial fluctuations smears out the rapid variations in the fully classical results. Note, however, that deep in the classical regime (large $\mathcal{S}$), these fluctuations could, in principle, be made arbitrary small and the interferences should reappear.  

\subsection{Full quantum analysis}
We now go beyond the classical and semi-classical approaches of the previous subsection and analyze the evolution of the full quantum system defined by the Hamiltonian~(\ref{lmgham}). One of the main objections is to explore whether the interference structure found in the transition probabilities in the classical model survives also in the quantum problem. Before presenting the results we may note that there are some earlier studies of related problems, but none has discussed the unavoidable interferences appearing in this model at a mean-field level. More precisely, driving the ferromagnetic LMG model through its critical point was analyzed in Refs.~\cite{lmgcrit}, and it was found that the non-adiabatic corrections obey a power-law dependence of the sweep velocity $\lambda$. A similar behaviour was also demonstrated in the {\it Tavis-Cummings model} describing $N$ spin-1/2 particles collectively interacting with a single boson mode~\cite{altland}. Also the LZ problem of the parity-broken LMG model has been considered~\cite{lmglz}. 

\begin{figure}[h]
\centerline{\includegraphics[width=8cm]{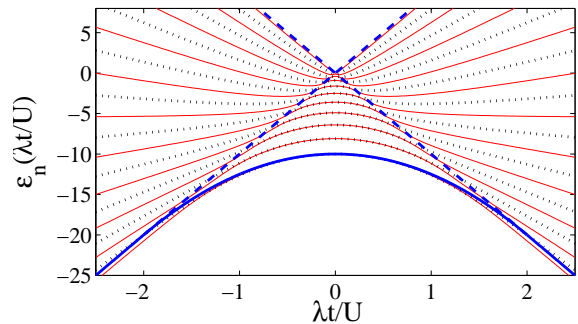}}
\caption{The spectrum of the LMG Hamiltonian $\hat{H}_\mathrm{LMG}$ for various $\lambda t/U$. The spin $\mathcal{S}=10$. The dotted black and solid red curves correspond to the two different parity solutions. The blue solid and dashed curves give the classical stable and unstable solutions respectively. The splitting and recombination of the classical solutions show the transcritical bifurcations.} \label{fig3}
\end{figure}

The eigenenergies $\varepsilon_n$ of $\hat{H}_\mathrm{LMG}$ are displayed in Fig.~\ref{fig3}. In the thermodynamic limit, the critical points are at $\lambda t/U=\pm2$ for which the two parity states become degenerate. Since the spectrum is symmetric with respect to $\lambda t/U=0$ it follows that the spectrum of the anti-ferromagnetic LMG is simply $-\varepsilon_n$. This demonstrates the fact that the anti-ferromagnetic LMG is not critical but hosts a first order quantum phase transition. 

When the initial ground state evolves it passes through a seam of level crossings starting at $t\approx-2U/\lambda$ and continuous until $t=2U/\lambda$. Thus, the system realizes a multi-channel Landau-Zener-St\"uckelberg interferometer. We note that this multi-level LZ crossings cannot, however, be described by the LZ {\it bow-tie model}~\cite{bt}. Interferences between the different paths (adiabatic states) could lead to final populations divided among the different diabatic states. In order to compare the amount of excitations in the present system to the LZ formula we introduce the projectors $\hat{P}_n(t)=|\psi_n\rangle\langle\psi_n|$, where $|\psi_n\rangle$ is the $n$'th instantaneous eigenstate of $\hat{H}_\mathrm{LMG}$, and define the excitation fraction as
\begin{equation}
P_\mathrm{ex}=\lim_{t\rightarrow\infty}\frac{1}{2\mathcal{S}+1}\sum_{n=0}^{2\mathcal{S}}n\langle\psi(t)|\hat{P}_{n+1}(t)|\psi(t)\rangle.
\end{equation}
Here, $|\psi(t)\rangle$ is the solution of the full time-dependent problem.
Thus, $P_\mathrm{ex}$ measures the amount of non-adiabatic excitations; $P_\mathrm{ex}=0$ corresponds to the case when only the ground state is populated while $P_\mathrm{ex}=1$ is the opposite limit of a maximally excited system. Note that $P_\mathrm{ex}$ is the mean of the final (scaled) distribution $P(n)$ of population of the various states $|\psi_n\rangle$. In the asymptotic limit $t\rightarrow+\infty$, when the diabatic and adiabatic states coincide, $\langle\hat{S}_z\rangle=\mathcal{S}(2P_\mathrm{ex}-1)$. To fully characterize the final distribution one would need all moments $\Delta^{(k)}n=\sum_nn^kP(n)$. Of particular interest is the Mandel $Q$-parameter~\cite{mandel} 
\begin{equation}\label{qpara}
Q=\frac{\Delta^{(2)}n-\left(\Delta^{(1)}n\right)^2}{\Delta^{(1)}n}-1
\end{equation}
which says whether the distribution $P(n)$ is sub- ($Q<0$) or super-Poissonian ($Q>0$). 

\begin{figure}[h]
\centerline{\includegraphics[width=8cm]{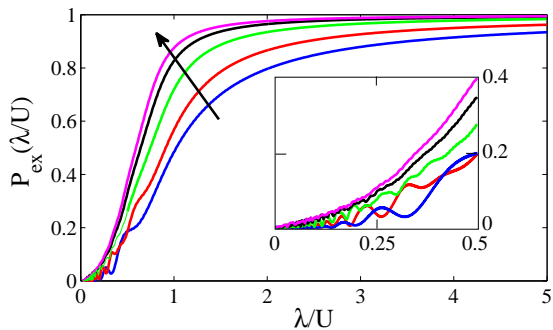}}
\caption{The average (scaled) number of excitations created during the LZ sweep for different spins $\mathcal{S}$: following the arrow 5 (blue), 12 (red), 24 (green), 50 (black), and 74 (magenta). In the adiabatic regime (inset) $P_\mathrm{ex}\sim\lambda^2$.} \label{fig4}
\end{figure}
 
The full time-dependent problem has been integrated from $t_i=-200$ to $t_f=+200$. We consider various spins $\mathcal{S}$ and sweep velocities $\lambda$. The results for $P_\mathrm{ex}$ are shown in Fig.~\ref{fig4}. We see in the figure that for growing spin $\mathcal{S}$ the system becomes more excited which can be understood from the increased density of states. Indeed, this is a result deriving from the critical slowing down mechanism in the vicinities of critical points. In the adiabatic and in the intermediate regimes we in particular find (numerically) that $P_\mathrm{ex}\sim\sqrt{\mathcal{S}}$.

In the adiabatic regime different power-law dependences $P_\mathrm{ex}\sim\lambda^\nu$ have been established in various types of LZ models; $\nu=3/4$ for parity-broken LMG model~\cite{wu}, $\nu=1$ for the Tavis-Cummings model~\cite{altland}, and $\nu=1/3$ (quantum regime) or $\nu=1$ (semi-classical regime) for a many-body fermionic model related to the Tavis-Cummings one~\cite{fleischhauer}. For a sweep through one of the critical point of the parity LMG it was found that $\nu=2$ deep in the adiabatic regime and $\nu=3/2$ in the intermediate regime~\cite{lmgcrit}. Such a dynamical situation is different from a full LZ sweep taken in this work where Landau-Zener-St\"uckelberg interferences can alter the excitations. Nevertheless, one may expect similar power-law dependences and this is indeed also the case as has been verified numerically. Thus, for small sweep velocities $\lambda$, $P_\mathrm{ex}\sim\lambda^2$ ({\it i.e.} $\nu=2$), and for the regime where breakdown of adiabaticity considerably sets in $P_\mathrm{ex}\sim\lambda^{3/2}$ ({\it i.e.} $\nu=3/2$) and finally in the diabatic regime we recover an exponential dependence $P_\mathrm{ex}\sim\left[1-\exp\left(-\kappa/\lambda\right)\right]$ for some $\lambda$-independent constant $\kappa$ (which is however $N$-dependent). Note that the corresponding quantum and semi-classical models display different scaling in the adiabatic regime. The inset of Fig.~\ref{fig4} display the excitations in the adiabatic regime, and we can hint that for large spins the exponential $\nu$ is actually smaller than 2 in the limit $\lambda\rightarrow0$ which could explain the discrepancy between the quantum and classical results; we can only expect agreement in the classical limit $\mathcal{S}\rightarrow\infty$.  

We now return to the question raised in the beginning of this section, {\it i.e.} will the interference phenomenon discussed in the classical model survive also in the quantum case? Clearly, from Fig.~\ref{fig4} we see that the classical oscillations are absent in the quantum simulations. One may say that this is of no surprise since the oscillations were already gone in the TWA result. However, as already pointed out, the limit $\mathcal{S}\rightarrow\infty$ should reproduce the classical results - the oscillations should appear for large enough spin values. This is certainly true for the TWA case at finite times since then the quantum uncertainty can be made vanishingly small. In the quantum case, for spins as large as $\mathcal{S}=500$ we have not been able to see any signatures of the classical oscillations. Furthermore, we see that the semi-classical TWA evolution is in general more adiabatic compared to the quantum one for spins $\mathcal{S}>20$. The difference between the quantum and classical results should be ascribed the multi-level Landau-Zener-St\"uckelberg interferences. We expect that such interferences should generate large fluctuations in the distribution $P(n)$. If this is the case one should find large Mandel $Q$-parameters as $\mathcal{S}$ grows. In the quasi-adiabatic and intermediate regimes we have numerically found that $Q\sim \mathcal{S}$, while from Fig.~\ref{fig4} we can extract that in the corresponding regime $P_\mathrm{ex}\sim\sqrt{\mathcal{S}}$. So the fluctuations relative to the non-adiabatic excitations in the system $Q/P_\mathrm{ex}\sim\sqrt{\mathcal{S}}$ which agrees with the findings of~\cite{altland}. Thus, as $\mathcal{S}$ is increased a larger number of final diabatic/adiabatic states will become populated.

\section{Conclusions}\label{seccon}
The LZ problem of a LMG model where the transition is driven by particle interaction was studied. At the mean-field level it was demonstrated that interference can drastically affect the transition probabilities throughout the different parameter regimes, and most surprisingly also deep in the adiabatic regime. At the quantum level, on the other hand, quantum fluctuations tend to `dephase' the dynamics such that rapid oscillations in the transition probabilities are suppressed. The mean-field and quantum results predicted different scaling behaviour of the transition probabilities, but this could arise from the finite system sizes used for the quantum simulations. 

We end by suggesting one possible system described by the parity LMG. By loading ultracold atoms into the first excited states of a two-dimensional anisotropic anharmonic trap, the quasi degeneracy of these states results in an effective spin-1/2 structure of the atoms~\cite{fernanda1}. The $\hat{S}_x^2$-term derives from the atom-atom interaction~\cite{fernanda1}. The anharmonicity of the trap prevents the atoms to decay into other energy states and by tuning the two trap frequencies the LZ sweep is realizable.

\begin{acknowledgments}
The author acknowledges Fernanda Pinheiro for discussions, and VR (Vetenskapsr\aa det) for financial help.
\end{acknowledgments}


\begin{thebibliography}{999}

\bibitem{lz} L. D. Landau, Phys. Z. Sowjetunion {\bf 2}, 46 (1932); 

\bibitem{zener} G. Zener, Proc. R. Soc. London Ser. A {\bf 137}, 696 (1932).

\bibitem{threelz1} O. Atabek, R. Lefebvre, and M. Jacon, J. Chem. Phys. {\bf 81}, 3874 (1984); C. E. Carroll and F. T. Hioe, J. Phys. A: Math. Gen. {\bf 19}, 2061 (1986); V. L. Pokrovsky and N. A. Sinitsyn, Phys. Rev. B {\bf 65}, 153105 (2002); J. Larson, Phys. Rev. A {\bf 73}, 013823 (2006).

\bibitem{bowtie} Y. N. Demkov and V. N. Ostrovsky, J. Phys. B: At. Mol. Opt. Phys. {\bf 34}, 2419 (2001).

\bibitem{julienne} V. A. Yurovsky, A. Ben-Reuven, and P. S. Julienne, Phys. Rev. A {\bf 65}, 043607 (2002).

\bibitem{altland} A. Altland {\it et al.}, Phys. Rev. A {\bf 79}, 042703 (2009).

\bibitem{fleischhauer} I. Tikhonenkov, E. Pazy, Y. B. Band, M. Fleischhauer, and A. Vardi, Phys. Rev. A {\bf 73}, 043605 (2006).

\bibitem{lmgcrit} P. Solinas, P. Ribeiro, and R. Mosseri, Phys. Rev. A {\bf 78}, 052329 (2008); T. Caneva, R. Fazio, and G. E. Santoro, Phys. Rev. B {\bf 78}, 104426 (2008); A. P. Itin and P. T\"orm\"a, arXiv:0901.4778.

\bibitem{orso} C. Kasztelan {\it et al.}, Phys. Rev. Lett. {\bf 106}, 155302 (2011).

\bibitem{LZoptlat} Y. Qian, M. Gong, and C. Zhang, Phys. Rev. A {\bf 87}, 013636 (2013).

\bibitem{nlz} B. Wu and Q. Niu, Phys. Rev. A {\bf 61}, 023402 (2000); O. Zobay and B. M. Garraway, Phys. Rev. A {\bf 61}, 033603 (2000).

\bibitem{wu} J. Liu {\it et al.}, Phys. Rev. A {\bf 66}, 023404 (2002).

\bibitem{lmglz} C. Lee, Phys. Rev. Lett. {\bf 97}, 150402 (2006); D. Witthaut, E. M. Graefe, and H. J. Korsch, Phys. Rev. A {\bf 73}, 063609 (2006); K. Smith-Mannschott {\it et al.}, Phys. Rev. Lett. {\bf 102}, 230401 (2009).

\bibitem{lmg2} R. G. Unanyan and M. Fleischhauer, Phys. Rev. Lett. {\bf 90}, 133601 (2003).

\bibitem{lmg} H. J. Lipkin, N. Meshkov, and A. J. Glick, Nucl. Phys. {\bf 62}, 188 (1965).


\bibitem{smel} M. S. Child, {\it Molecular Collision Theory}, (Academic Press, London, 1974); A. P. Kazantsev, G. I. Surdutovich, D. O. Chudesnikov, and V. P. Yakovlev, J. Phys. B: At. Mol. Phys. {\bf 18}, 2619 (1985).

\bibitem{stokes} C. M. Bender and S. A. Orszag, {\it Advance Mathematical Methods for Scientists and Engineers}, (McGraww-Hill, New York, 1978).

\bibitem{kalle} K.-A. Suominen, {\it The Landau-Zener Linear Crossing Problem and Non-Adiabatic Transitions}, Licentiate Thesis.

\bibitem{stuck} S. N. Shevchenko, S. Ashab, and F. Nori, Phys. Rep. {\bf 492}, 1 (2010).

\bibitem{lmg1} A. Micheli, D. Jaksch, I. Cirac, and P. Zoller, Phys. Rev. A {\bf 67}, 013607 (2003).

\bibitem{lmg3} S. Morrison and A. S. Parkins, Phys. Rev. Lett. {\bf 100}, 040403 (2008); D. I. Tsomokos, S. Ashhab, and F. Nori, New J. Phys. {\bf 10}, 113020 (2008); G. Chen {\it et al.}, Opt. Express {\bf 17}, 19682 (2009); J. Larson, Eur. Phys. Lett. {\bf 90}, 54001 (2010).

\bibitem{sakurai} A. Auerback, {\it Interacting electrons and quantum magnetism}, (Springer Verlag, New York, 1998).

\bibitem{sachdev} S. Sachdev, {\it Quantum Phase Transitions}, (Cambridge University Press, Cambridge, 2011).

\bibitem{ll} L.D. Landau and E. M. Lifschitz, {\it Mechanics} (Pergamon, Oxford, 1977).

\bibitem{nlzexp} Y.-A. Chen {\it et al.}, Nat. Phys. \textbf{7}, 61 (2011).

\bibitem{twa} A. Polkovnikov, Annals Phys. {\bf 325}, 1790 (2010).

\bibitem{bt} V. N. Ostrovsky and H. Nakamura, J. Phys. A: Math. Gen. {\bf 30}, 6939 (1997); Y. N. Demkov and V. N. Ostrovsky, Phys. Rev. A {\bf 61}, 032705 (2000).

\bibitem{mandel} L. Mandel and E. Wolf, {\it Optical Coherence and Quantum Optics}, (Cambridge University Press, Cambridge, 1995).

\bibitem{fernanda1} F. Pinheiro, J.-P. Martikainen, and J. Larson, Phys. Rev. A {\bf 85}, 033638 (2012); F. Pinheiro, G. M. Bruun, J.-P. Martikainen, and J. Larson, Phys. Rev. Lett. {\bf 111}, 205302 (2013).
\bibitem{fernanda1} F. Pinheiro, J.-P. Martikainen, and J. Larson, Phys. Rev. A {\bf 85}, 033638 (2012); F. Pinheiro {\it et al.}, Phys. Rev. Lett. {\bf 111}, 205302 (2013).































\end{thebibliography}
\end{document}